\documentclass[epj]{svjour}
\usepackage{booktabs}
\usepackage{pstricks}

\usepackage{mathtools}
\usepackage[shortskips]{methtools}
\usepackage{dsfont}
\usepackage{slashed}
\usepackage[numbers, comma, sort&compress]{natbib}

\usepackage{amssymb,latexsym}
\usepackage[squaren,Gray]{SIunits}

\newcommand{\overbar}[1]{\mkern 1.5mu\overline{\mkern-1.5mu#1\mkern-1.5mu}\mkern 1.5mu}
\newcommand{\MSbar}{\overbar{\text{MS}}}
\newcommand{\tr}[1]{\operatorname{tr} \left\{ #1 \right\}}

\begin{document}%
%
%
\title{%
Pion distribution amplitude from Euclidean correlation functions
}
%
%
%
\author{%
Gunnar~S.~Bali		\inst{1,2}	\and
Vladimir~M.~Braun	\inst{1}	\and
Benjamin~Gl\"a{\ss}le	\inst{1}	\and
Meinulf~G\"ockeler	\inst{1}	\and
Michael~Gruber		\inst{1}	\and
Fabian~Hutzler		\inst{1}	\and
Piotr~Korcyl		\inst{1,3}	\and
Bernhard~Lang		\inst{1}	\and
Andreas~Sch\"afer	\inst{1}	\and
Philipp~Wein		\inst{1}\fnmsep\thanks{\email{philipp.wein@physik.uni-r.de}}	\and
Jian-Hui~Zhang		\inst{1}
}%
%
%
\institute{%
Institut f{\"u}r Theoretische Physik, Universit{\"a}t Regensburg,\\%
Universit{\"a}tsstra{\ss}e 31, 93040 Regensburg, Germany
\and
Department of Theoretical Physics, Tata Institute of Fundamental Research,\\%
Homi Bhabha Road, Mumbai 400005, India
\and
Marian Smoluchowski Institute of Physics, Jagiellonian University,\\%
ul.\ {\L}ojasiewicza 11, 30-348 Krak\'ow, Poland
}%
%
%
\abstract{%
Following the proposal in~\cite{Braun:2007wv}, we study the feasibility to calculate the pion distribution amplitude (DA) from suitably chosen Euclidean correlation functions at large momentum. In our lattice study we employ the novel momentum smearing technique~\cite{Bali:2016lva,Bali:2017ude}. This approach is complementary to the calculations of the lowest moments of the DA using the Wilson operator product expansion and avoids mixing with lower dimensional local operators on the lattice. The theoretical status of this method is similar to that of quasi-distributions~\cite{Ji:2013dva}, that have recently been used in~\cite{Zhang:2017bzy} to estimate the twist two pion DA. The similarities and differences between these two techniques are highlighted.%
}%
%
%
\maketitle
%
%
\section{Introduction}
In recent years there has been increasing interest in the possibility to determine parton distribution functions from Euclidean correlation functions, bypassing Wilson's operator product expansion. The general scheme of such calculations is to consider a product of suitable local currents at a spacelike separation $z$, sandwiched between hadronic states,%
\begin{align} \label{matrix_element}
  \langle H| \,\big(\bar q\Gamma_1 Q\big)(z/2)\,\big(\bar Q\Gamma_2 q\big)(-z/2)\,|H'\rangle\,, 
\end{align}%
and match the lattice calculation of this quantity to the perturbative expansion in terms of collinear parton distributions%
\begin{align}
    \langle H| \bar q(n) \Gamma q(-n) |H'\rangle\,, \qquad n^2=0\,.
\end{align}%
The existing concrete proposals differ mainly in the choice of the $Q$-field. This can be chosen as an auxiliary scalar in the fundamental representation of the color group~\cite{Aglietti:1998ur,Abada:2001if}, or as an (auxiliary) heavy~\cite{Detmold:2005gg} or light~\cite{Braun:2007wv} quark. Another suggestion~\cite{Ji:2013dva} is to replace the $Q$-field propagator by a Wilson line connecting $\bar q(z/2)$ and $q(-z/2)$. This last proposal received the most attention, despite added complications due to the renormalization of the Wilson line~\cite{Craigie:1980qs,Dorn:1986dt}, see Refs.~\cite{Ishikawa:2016znu,Ishikawa:2017jtf,Rossi:2017muf,Ji:2017rah,Ji:2017oey,Orginos:2017kos,Alexandrou:2017huk,Green:2017xeu} for recent discussions, the reason being that it allows for a more direct momentum space interpretation in the framework of the large-momentum effective theory~\cite{Ji:2017rah,Ji:2014gla} (LaMET). The corresponding correlation functions, transformed into the longitudinal momentum fraction representation, have become known as quasi-parton-distributions~\cite{Ji:2013dva}. While quasi parton-distributions are certainly interesting objects, it was already pointed out in Refs.~\cite{Braun:2007wv,Braun:1994jq} that position space correlation functions (or ``lattice cross sections'', in the terminology of~\cite{Ma:2014jla,Ma:2014jga,Ma:2017pxb}) contain the complete information on parton distributions. In all cases the functions calculated on the lattice (for early work, see also~\cite{Liu:1993cv}) are related to parton distributions by means of QCD factorization in the continuum, which can be done both in position and momentum space. A position space analysis naturally leads to the concept of Ioffe-time distributions~\cite{Orginos:2017kos,Braun:1994jq,Radyushkin:2017cyf,Radyushkin:2017gjd}. We emphasize that all the above suggestions are equivalent, and their relative virtue will be determined by the possibility to control lattice artifacts and other systematic uncertainties.\par
In this work we study the simplest function of this kind, the pion distribution amplitude (DA), using the technique suggested in Ref.~\cite{Braun:2007wv}, i.e., we use a light quark ($Q=q$) in Eq.~\eqref{matrix_element} and perform the analysis directly in position space. The same DA has recently been studied using the quasi-distribution approach in Ref.~\cite{Zhang:2017bzy}. We consider a correlation function of renormalized scalar and pseudoscalar operators at equal times%
\begin{align} \label{T_SP}
T(p \cdot z,z^2)& = \langle \pi^0(\mathbf p ) | \big[\bar u\, q\big](z/2) 
\, \big[\bar q \gamma_5 u\big](-z/2) | 0 \rangle\,, 
\end{align}%
where $\bar q(z)$ creates a light quark field of hypothetical flavor $q\neq u,d$ and square brackets $[\mathcal {O}]$ denote operator renormalization in the $\MSbar$ scheme. In what follows, we fix the renormalization scale to the ``kinematic'' scale in the correlator, $\mu_R = 2/\sqrt{-z^2}$ (cf.~\cite{Braun:2007wv,Chetyrkin:2010dx}). The correlation function~\eqref{T_SP} can be calculated on the lattice as a function of two variables, $p\cdot z = - \mathbf{p} \cdot\mathbf{z} $ and $z^2 = -\mathbf{z}^2$. Here and below we use boldface letters for spatial 3-vectors.\par%
We restrict ourselves to sufficiently small distances, $|\mathbf{z}|/2 < \unit{1}{\power{\giga\electronvolt}{-1}}$, such that the same correlation function can be calculated in continuum perturbation theory in terms of the pion DA using standard QCD factorization techniques. The result reads%
\newlength\templ%
\settoheight{\templ}{\includegraphics[width=\columnwidth,clip]{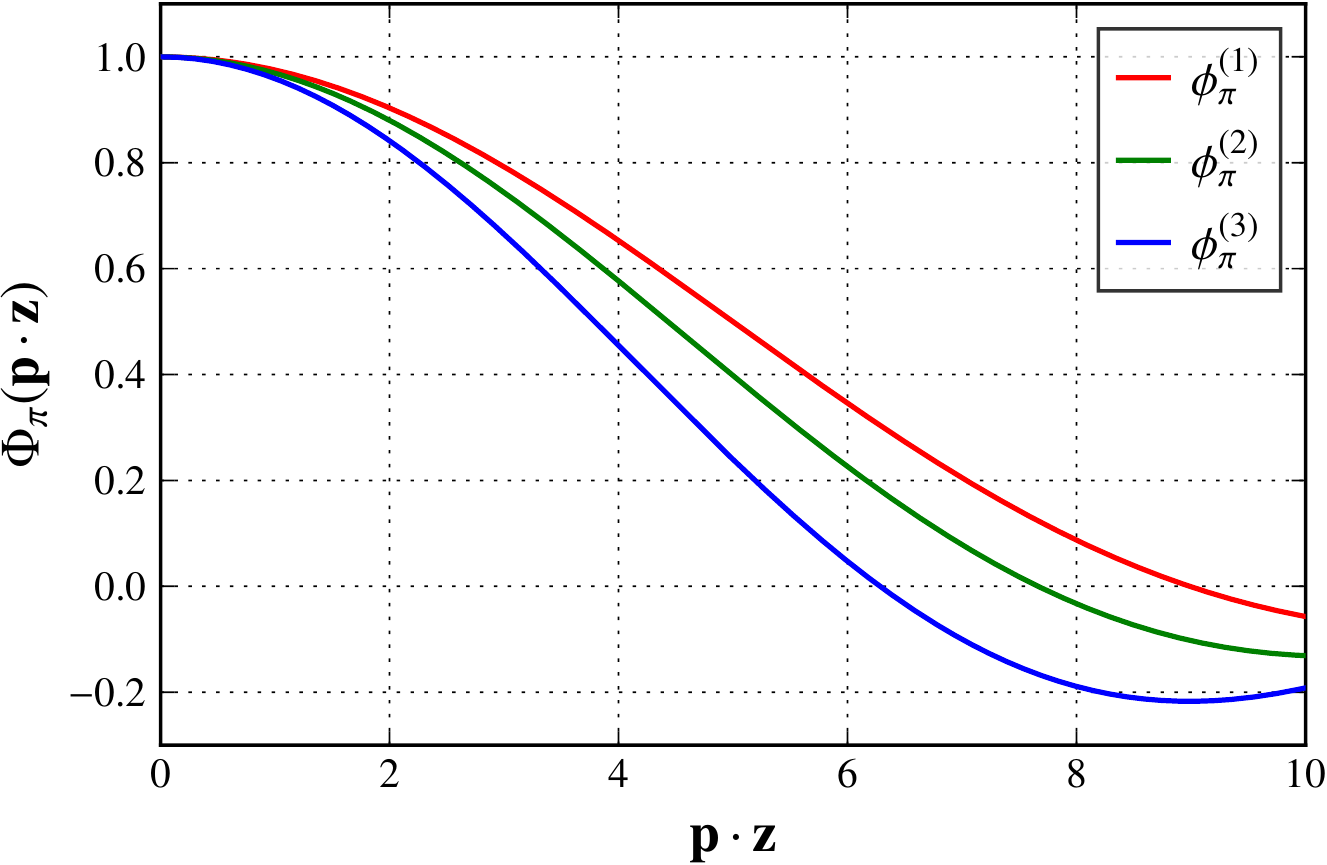}}%
\begin{figure}[t]%
\begin{minipage}[t][\templ]{\columnwidth}%
\centering
\includegraphics[width=\columnwidth,clip]{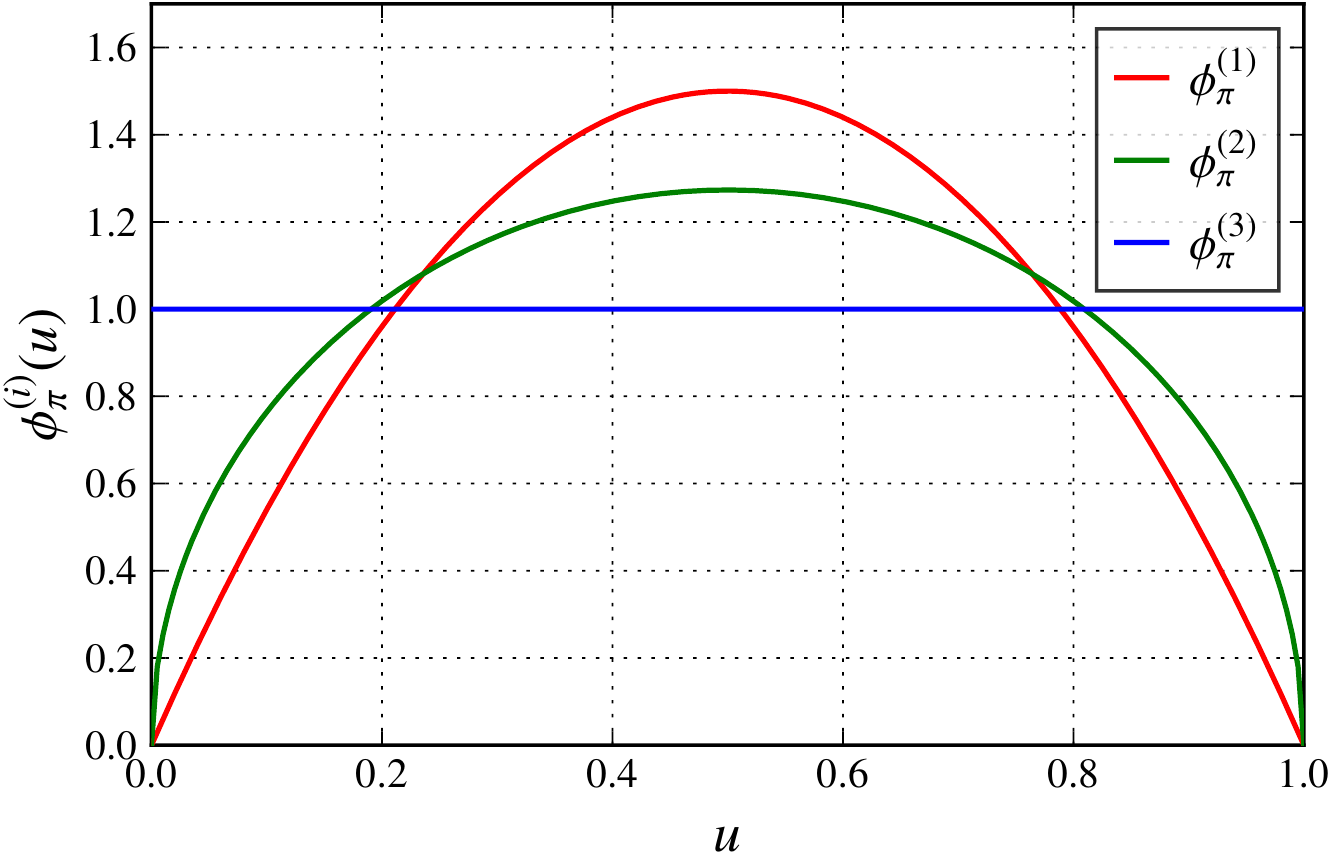}%
\end{minipage}%
\caption{\protect\rule{0sp}{\topskip}\label{fig_pionDA}Plot of the three models for the pion distribution \mbox{amplitude} given in Eq.~\eqref{models}.}
\end{figure}%
\begin{align}
\label{T_SP_LO}
   T(p \cdot z,z^2) &= F_\pi \frac{p\cdot z}{2\pi^2 z^4}  \Phi^{\rm SP}_\pi(p\cdot z,z^2)\,, 
\end{align}
where $\Phi^{\rm SP}_\pi=\Phi_\pi + \mathcal O(\alpha_s) + \text{higher twist}$ (the various corrections will be discussed later), with
\begin{align}
 \Phi_\pi(p\cdot z) &= \int_0^1\!du\, e^{ i(u-1/2)(p\cdot z)} \phi_\pi(u)\,.
\label{Phi_pi}
\end{align}%
$F_\pi\approx\unit{93}{\mega\electronvolt}$ is the pion decay constant, the variable~$u$ corresponds to the quark momentum fraction and $\phi_\pi(u)$ is the (leading-twist) pion DA. The integral of the pion DA is normalized to unity, $\int_0^1\!du\,\phi_\pi(u)=1,$ and its shape has been hotly debated for more than 30~years. This discussion has been reinvigorated by the strong scaling violation in the $\pi\gamma^*\gamma$ form factor observed by the BABAR~\cite{Aubert:2009mc} and, to a lesser extent, the BELLE~\cite{Uehara:2012ag} collaboration, which is difficult to explain unless the pion DA exhibits strong enhancements near the end points, see, e.g., Refs.~\cite{Agaev:2010aq,Mikhailov:2016klg,Horn:2016rip} for a review and further references. For illustrative purposes we consider three models:%
\begin{align}
\label{models}
\phi_\pi^{(1)} (u) &= 6 u(1-u)\,,
\notag\\
\phi_\pi^{(2)} (u) &=  \frac{8}{\pi}\sqrt{u(1-u)}\,,
\notag\\
\phi_\pi^{(3)} (u) &=  1\,,
\end{align}%
\begin{figure}[t]%
\centering
\includegraphics[width=\columnwidth,clip]{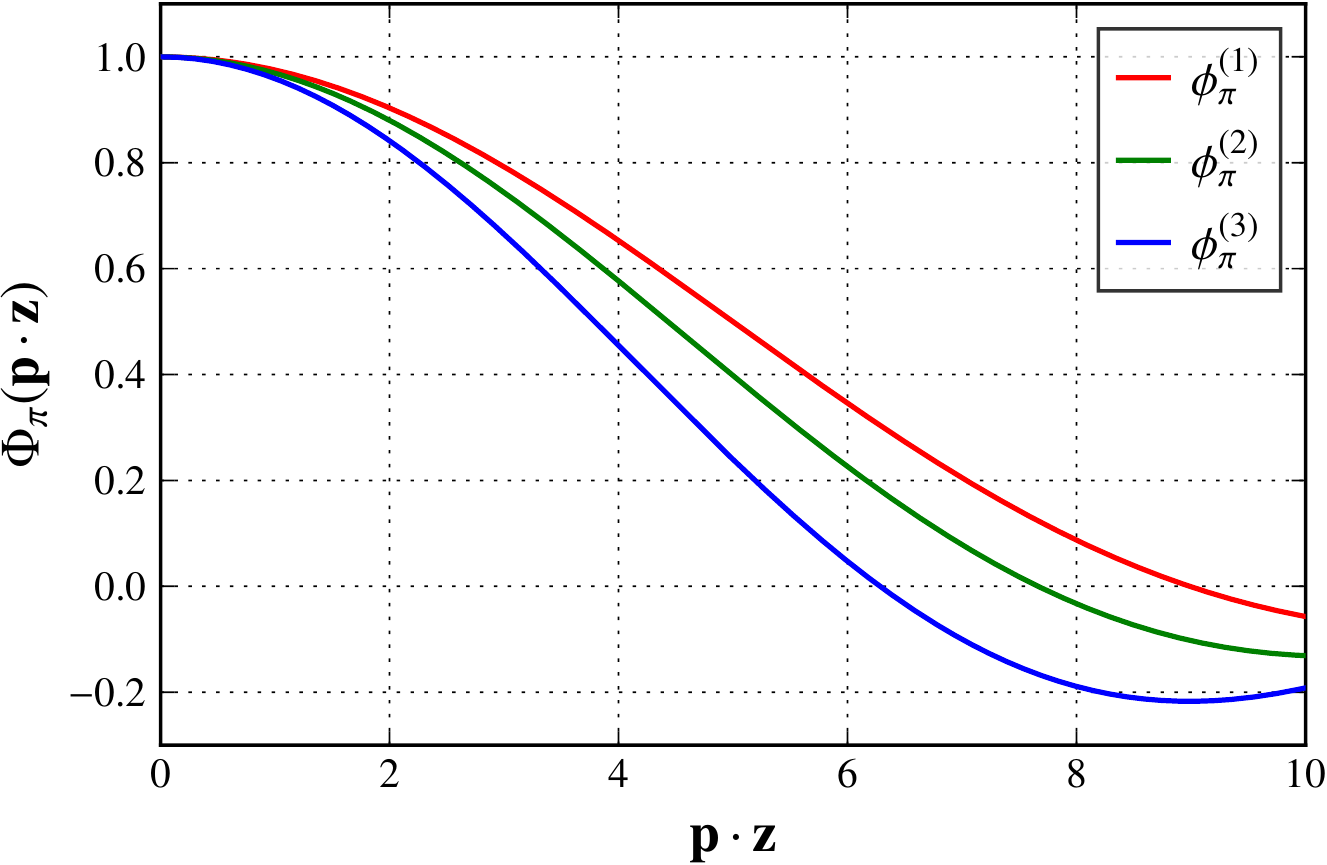}%
\caption{\label{fig_pionDA_position_space}The position space pion DA $\Phi_\pi(p\cdot z)$~[cf.\ Eq.~\eqref{Phi_pi}] for the three models in Eq.~\eqref{models}.}
\end{figure}%
at the reference scale $\mu_0=\unit{1}{\giga\electronvolt}$. These models and the corresponding Fourier-transformed position space DAs $\Phi_\pi(p\cdot z)$, defined in Eq.~\eqref{Phi_pi}, are plotted in Figs.~\ref{fig_pionDA} and~\ref{fig_pionDA_position_space}, respectively. Measuring the correlation function~\eqref{T_SP} on the lattice for a range of values of $p\cdot z = - \mathbf{p} \cdot\mathbf{z} $ and $z^2 = -\mathbf{z}^2$ gives access to the pion DA in position space~\eqref{Phi_pi} that contains the full information on the longitudinal momentum fraction distribution.\par%
The main difference of our technique~\cite{Braun:2007wv} to the approach of Ref.~\cite{Zhang:2017bzy} is that the smallness of higher twist and perturbative corrections (for \emph{arbitrary} pion momentum) is guaranteed by keeping the distance $|\mathbf{z}|$ between the currents sufficiently small. A large pion momentum is needed not in order to suppress the corrections, but because it provides the necessary lever arm in the dimensionless variable $\mathbf{p} \cdot\mathbf{z}$ that is mandatory to distinguish between pion DAs of different shape, see Fig.~\ref{fig_pionDA_position_space}.\par%
In contrast, in the LaMET-based approach of~\cite{Zhang:2017bzy} formally a Fourier transform over all values of $z$ is taken, and smallness of perturbative and higher twist corrections is achieved indirectly by considering the asymptotic expansion of the amplitude at large values of the pion momentum for a fixed quark momentum fraction $u$, which is the Fourier conjugate of $\mathbf{p} \cdot\mathbf{z}$. Thus $|\mathbf{p}|\to \infty$ implies that the integration region in the Fourier integral shrinks to  $ |\mathbf{z}|\sim 1/|\mathbf{p}| \to 0$.\par%
Another difference is that in Ref.~\cite{Zhang:2017bzy} a Wilson line is used to connect the quark and the antiquark, whereas in this study we use a light-quark propagator~\cite{Braun:2007wv}. To tree-level accuracy the difference in the corresponding coordinate space expressions is simply a different coefficient function in Eq.~\eqref{T_SP_LO}. While taking the Wilson line not along a lattice axis introduces additional difficulties, e.g., concerning renormalization, the separation $\mathbf z$ can be chosen arbitrarily without problems when a light-quark propagator is used. We consider this possibility as an advantage of our calculation because we have found discretization errors to be largest if $\mathbf z$ lies along a lattice axis. Also the renormalization of the lattice correlator is greatly simplified when one works with a light-quark propagator (for recent progress regarding the Wilson line approach see~\cite{Orginos:2017kos,Alexandrou:2017huk,Green:2017xeu}).\par%
Note that we suggest to match the lattice matrix element with the pQCD factorization expression directly in coordinate space. This has the advantage that the lattice data can be directly confronted with the theory since perturbative predictions based on model parametrizations of the DAs can easily be transformed to position space.\par%
The whole program naturally splits into two parts --- the lattice calculation where all usual extrapolations/limits have to be taken and the pQCD factorization in terms of the pion DA in the continuum. Our presentation is structured accordingly.%
\section{QCD factorization}%
The complete QCD expression for the correlation function~\eqref{T_SP} can be written as%
\begin{flalign}%
&\mathrlap{T(p \cdot z,z^2) ={}}
\\&&\mathllap{{}= F_\pi \frac{p\cdot z}{2\pi^2 z^4} \int\limits_0^{\smash{1}}\!du\,e^{i(u-1/2)(p\cdot z)} \, H(u,z^2,\mu) \phi_\pi(u,\mu) + T^{\rm HT}\,,}&\notag
\end{flalign}%
where $H(u,z^2,\mu) = 1 + \mathcal{O}(\alpha_s)$ is a short distance coefficient function that can be evaluated perturbatively, $\mu$ is the factorization scale, and $T^{\rm HT}$ stands for power-suppressed (in $z^2$) contributions that can be calculated in terms of the pion DAs of higher twist~\cite{Braun:1989iv,Ball:2006wn}.\par%
The factorization scale dependence of the pion DA is considerably simplified by using the expansion%
\begin{align}
\phi_\pi(u,\mu) &= 6u(1-u) \sum\limits_{n=0}^{\infty} a^\pi_n(\mu) C^{3/2}_n(2u-1) \,,
\end{align}%
where the $C^{3/2}_n(x)$ are Gegenbauer polynomials. The $n=0$ coefficient is fixed to unity, $a^\pi_0=1$, by the normalization condition and the remaining ones, $n=2,4,\ldots$, encode all relevant nonperturbative information on the DA. They have to be defined at a certain reference scale $\mu_0$ (a common choice is $\mu_0=\unit{1}{\giga\electronvolt}$) and evolved to the scale of the process. The corresponding mixing matrices are known in analytic form to two-loop accuracy~\cite{Mikhailov:1985cm,Mueller:1993hg} and numerically for the first few moments to three-loop accuracy~\cite{Braun:2017cih}.\par%
Using this expansion, the leading-twist (LT), i.e., twist two, contribution to the correlation function can be written as%
\begin{align}
\label{nsum}
     T^{LT} &= F_\pi \frac{p\cdot z}{2\pi^2 z^4} \sum\limits_{n=0}^{\infty} H_n(p\cdot z,\mu)\, a^\pi_n(\mu)\,.
\end{align}%
Setting both the renormalization and factorization scales to $\mu\equiv2/\sqrt{-z^2}$ we obtain, to $\mathcal{O}(\alpha_s)$ accuracy,%
\begin{flalign} \label{Hn}
 H_n &= \biggl[ 1 + \frac{\alpha_s C_F}{4\pi}(7 \eta - 11)  \biggr]\, \mathcal F_n\bigl(\tfrac12  p \cdot z\bigr) &&\\& \quad 
  -\frac{\alpha_s C_F }{\pi} \! \int_0^1 \! ds\, 
  \mathcal F_n\bigl(\tfrac{s}{2} p \cdot z\bigr)\, \Biggl\{ (\eta - 4) \frac{\sin\bigl(\tfrac{\bar s}{2} p \cdot z\bigr)}{p \cdot z} 
\notag\\ &&& 
  \mathllap{+ \biggl( (\eta - 2) \biggl[\frac{s}{\bar s}\biggr]_+ + \biggl[\frac{\ln(\bar s)}{\bar s}\biggr]_+ \biggr) \cos\bigl(\tfrac{\bar s}{2} p \cdot z\bigr) \Biggr\}\,,}\notag
\end{flalign}%
where $C_F=\frac43$, $\eta=1+2\gamma_E$, $\bar s=1-s$, and%
\begin{align}
 \mathcal F_n(\rho) &= \frac34 i^n \sqrt{2 \pi} (n+1) (n+2) \rho^{-\frac32} J_{n+\frac32}(\rho) \,.
\end{align}%
The plus prescription is defined as usual,%
\begin{align}
 \int\limits_0^1 ds \; f(s) \bigl[ g(s) \bigr]_+ \equiv   \int\limits_0^1 ds\; \bigl[f(s)-f(1)\bigr]  g(s) \,.
\end{align}%
The sum in~\eqref{nsum} converges very rapidly since%
\begin{align}
  \mathcal F_n(\rho) &\stackrel{\rho\to 0}{\simeq} 
\frac38 i^n \left(\frac{\rho}{2}\right)^n \frac{\sqrt{\pi}(n\!+\!1)(n\!+\!2)}{\Gamma(n+5/2)}\,,
\end{align}%
so that for finite $\rho \sim \frac12 p\cdot z$ only the first few Gegenbauer moments give a sizeable contribution, cf.~\cite{Braun:2007wv}.\par%
The leading higher twist contribution $\mathcal{O}(z^2)$ can be estimated using models for the twist~$4$ pion DAs discussed in Refs.~\cite{Braun:1989iv,Ball:2006wn}. For the case at hand these corrections are in general complex. We obtain for the real part%
\begin{flalign} 
& \operatorname{Re}\mathrlap{T^{\rm HT} =  
  - F_\pi \frac{p\cdot z}{8\pi^2 z^2 }\int \limits_0^1 du \, \cos[(u-1/2) p\cdot z] \biggl\{ 
 20 \delta^2_\pi u^2\bar u^2 }
\notag\\ &&& - m^2_\pi u\bar u + \frac12 m^2_\pi u^2\bar u^2
 \Big[14u\bar u -5 + 6a_2^\pi (3-10u\bar u)\Big] \biggr\} \,,
\end{flalign}%
where $\delta^2_\pi \simeq \unit{0.2}{\giga\electronvolt\squared}$ at the scale $\mu=\unit{1}{\giga\electronvolt}$~\cite{Braun:1989iv,Ball:2006wn}. The last two terms take into account the pion mass corrections and are rather small.\par%
We find that the higher twist correction for the scalar-pseudoscalar correlation function has the same sign as the leading-twist term, in contrast to the vector-vector correlation function considered in Ref.~\cite{Braun:2007wv}, in which case the higher twist correction has the opposite sign. Numerically, this correction turns out to be about $20\%$ of the leading twist contribution at $|\mathbf z|/2 \sim \unit{0.2}{\femto\meter} \simeq \unit{1}{\power{\giga\electronvolt}{-1}}$.%
\section{Lattice calculation}%
\subsection{Generalities}%
\begin{figure}[tb]%
\centering%
\includegraphics[width=\columnwidth,trim=0 -2 32 0, clip]{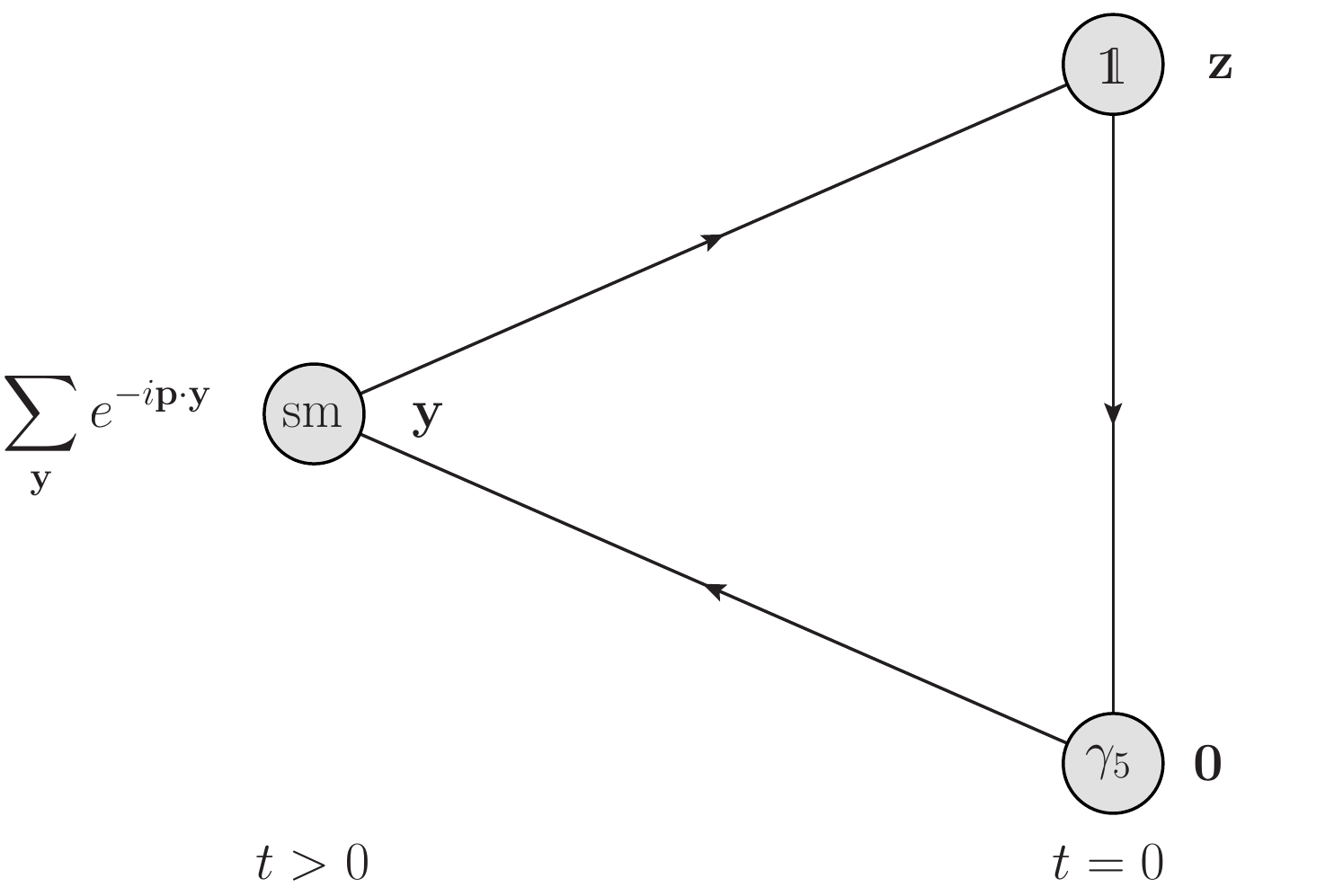}%
\caption{\label{fig_triangle}The relevant triangle diagram, where the two local currents are at $t=0$, while the smeared interpolating current for the pion with momentum $\mathbf p$ is situated at $t>0$.}
\end{figure}%
We wish to avoid the calculation of disconnected quark line diagrams, which are challenging in lattice simulations. This becomes possible by implementing an appropriate flavor structure of our currents. One may consider having a $\pi^0$ in the final state and $q=d$ in Eq.~\eqref{T_SP}. However, this matrix element vanishes identically due to isospin symmetry. Instead we pretend that the auxiliary quark field $q$ of Eq.~\eqref{T_SP} is a different, third flavor but for simplicity we keep it at the same mass $m_q=m_u=m_d$. This corresponds to our continuum QCD calculation.\par%
In the actual lattice calculation we determine the three-point function using the sequential source method. Therefore, the currents are situated at $\mathbf z$ and at (the chosen origin) $\mathbf 0$ and are afterwards ``shifted'' to the symmetric locations in Eq.~\eqref{T_SP} by multiplication with the appropriate phase.\par%
\subsection{Correlation functions}%
\begin{figure}[t]%
\centering
\newlength{\tempp}%
\settodepth{\tempp}{\small Points with a correction larger than $10\%$ (gray triangles) will be ignored in the analysis.}%
\includegraphics[width=\columnwidth]{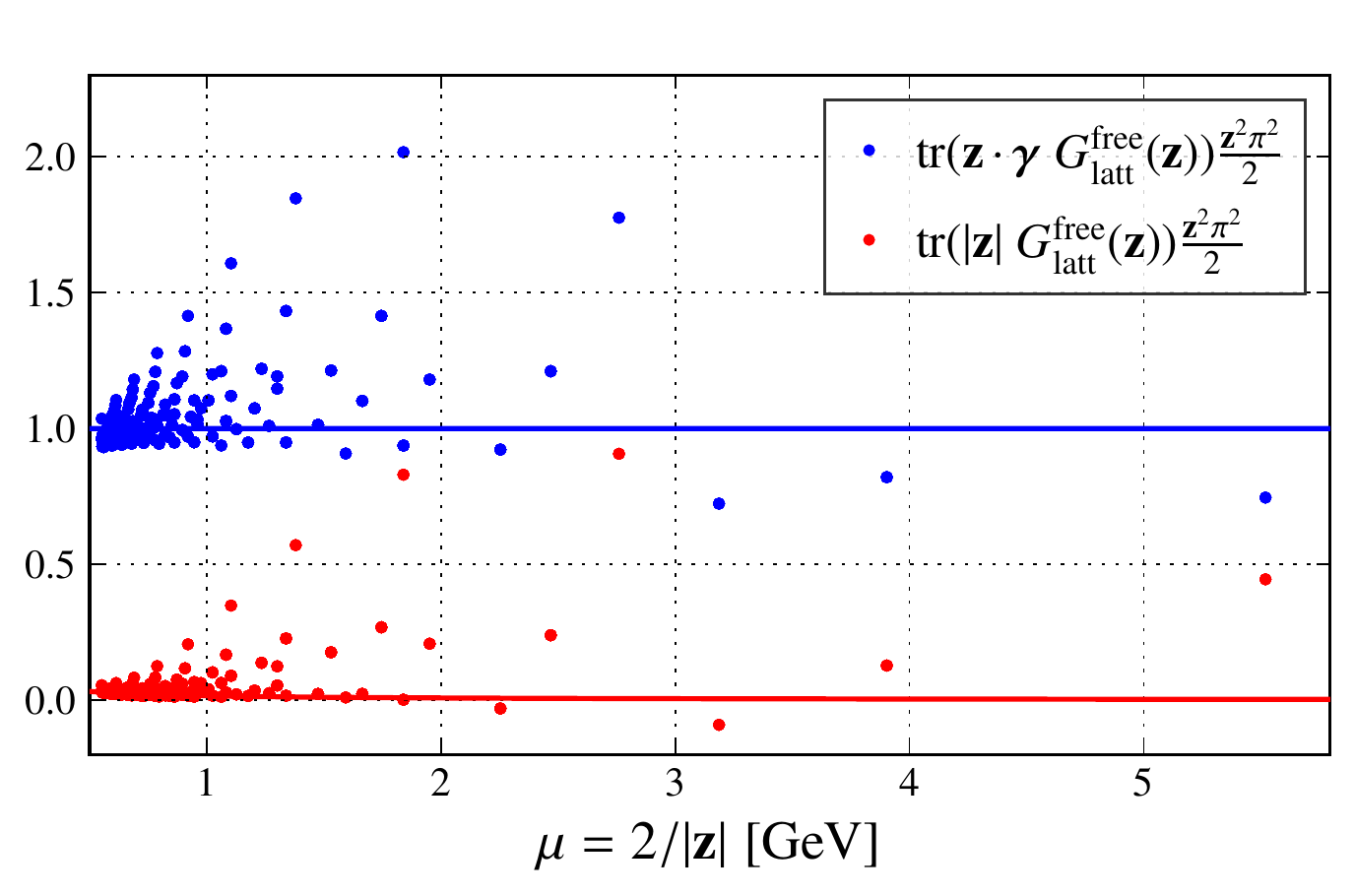}
\caption{\label{fig_disc}The (free field) discretization effects of the Wilson propagator compared to the continuum expectation for the different Dirac structures.\protect\rule[-\tempp]{0pt}{\tempp}}
\end{figure}%
\begin{figure}[t]%
\centering
\includegraphics[width=\columnwidth]{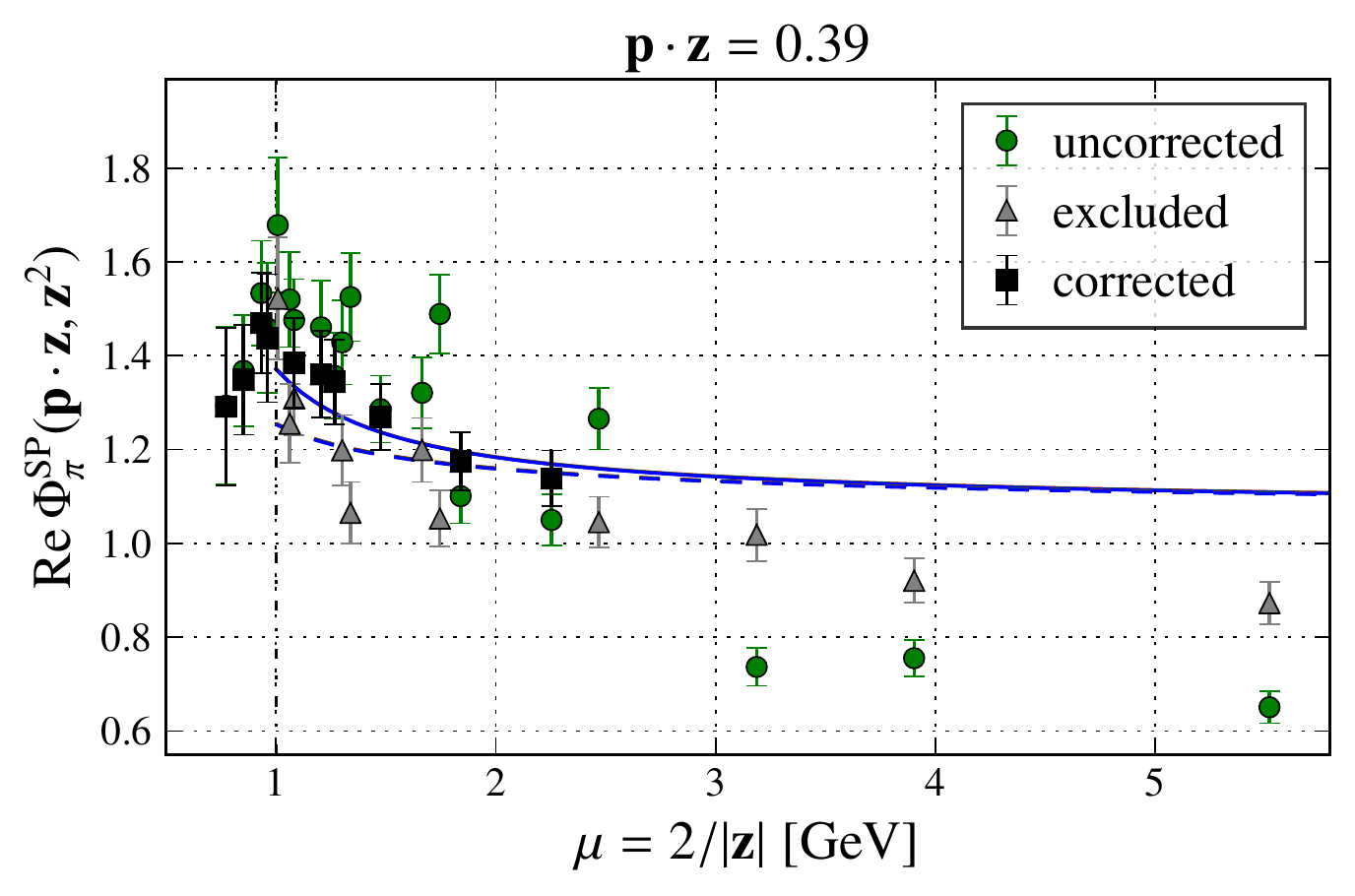}
\caption{\label{fig_disc_corr}Correction of discretization effects for the example of a fixed $\mathbf p \cdot \mathbf z=0.39$ and $|\mathbf p|=\unit{1.08}{\giga\electronvolt}$. Points with a correction larger than $10\%$ (gray triangles) will be ignored in the analysis.}
\end{figure}%
The remaining nontrivial part of the lattice simulation is the calculation of the connected triangle diagram depicted in Fig.~\ref{fig_triangle}. Introducing a phase matrix $\varphi_t$ that is diagonal in position space (with diagonal entries $(\varphi_t)_{\mathbf y \mathbf y}=e^{-i \mathbf p \cdot \mathbf y}$) and is nonzero only on time slice $t$, we can rewrite the three-point function as follows:%
\begin{align}%
C^{\text{3pt}}&= i   \Bigl\langle \tr{G(z,0) \gamma_5 \Bigl( G \Phi_{(\mathbf p)} \varphi_t \gamma_5 \Phi_{(-\mathbf p)} G\Bigr)(0,z) } \Bigr\rangle \nonumber \\
&\equiv i   \Bigl\langle \tr{S(z,0)^\dagger \gamma_5 G(z,0) } \Bigr\rangle  \,,
\end{align}%
where $z=(\mathbf z, 0)$, $G$ stands for the quark propagator, and%
\begin{align}%
 S=\gamma_5 \Bigl(G \Phi_{(\mathbf p)} \varphi_t \gamma_5 \Phi_{(-\mathbf p)} G \Bigr)^\dagger \gamma_5 =  G \Phi_{(-\mathbf p)} \varphi_t^\dagger \gamma_5 \Phi_{(\mathbf p)} G \,,
\end{align}%
is a sequential source. The momentum dependent smearing $\Phi_{(\mathbf p)}$ is performed as described in Ref.~\cite{Bali:2016lva}. We want to stress that in this situation the new momentum smearing technique is even more cost efficient than described in Ref.~\cite{Bali:2017ude}, since one needs a second inversion of the Dirac operator for each additional momentum anyway.\par%
The matrix element~\eqref{T_SP} can be obtained from $C^{\text{3pt}}$ by canceling the normalization factor describing the overlap of the smeared current with the pion state. The latter can be obtained, e.g., from the two-point function $C^{\text{2pt}}$ of a smeared current at the sink (at time $t$, as in the three-point function) and a local axialvector current at the source. Neglecting excited state contributions, one finds%
\begin{align}%
 \frac{T(p\cdot z,z^2)}{F_\pi} &= \frac{Z_S(\mu)Z_P(\mu)}{Z_A} \frac{C^{\text{3pt}}(\mathbf p,\mathbf z)}{C^{\text{2pt}}(\mathbf p)} E(\mathbf p) \,,
\end{align}%
where $Z_X$ is the renormalization factor of the local current $X$ with respect to the $\MSbar$ scheme~\cite{Gockeler:2010yr}, cf.\ Sect.~\ref{sect_results}.
\subsection{Taming discretization effects}\label{sect_disc_corr}%
\begin{figure*}[t]%
\centering%
\includegraphics[width=\columnwidth]{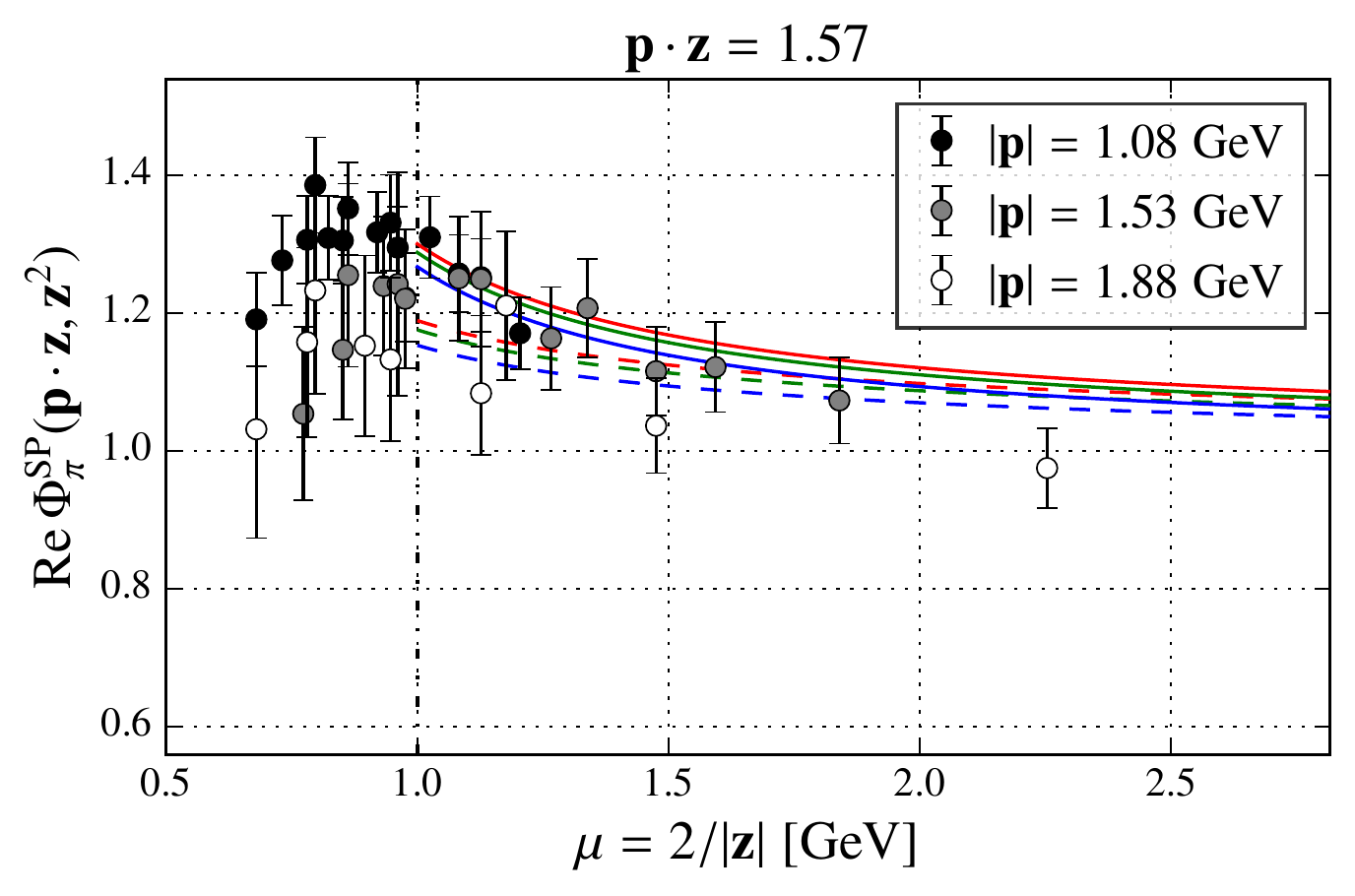}\hfill%
\includegraphics[width=\columnwidth]{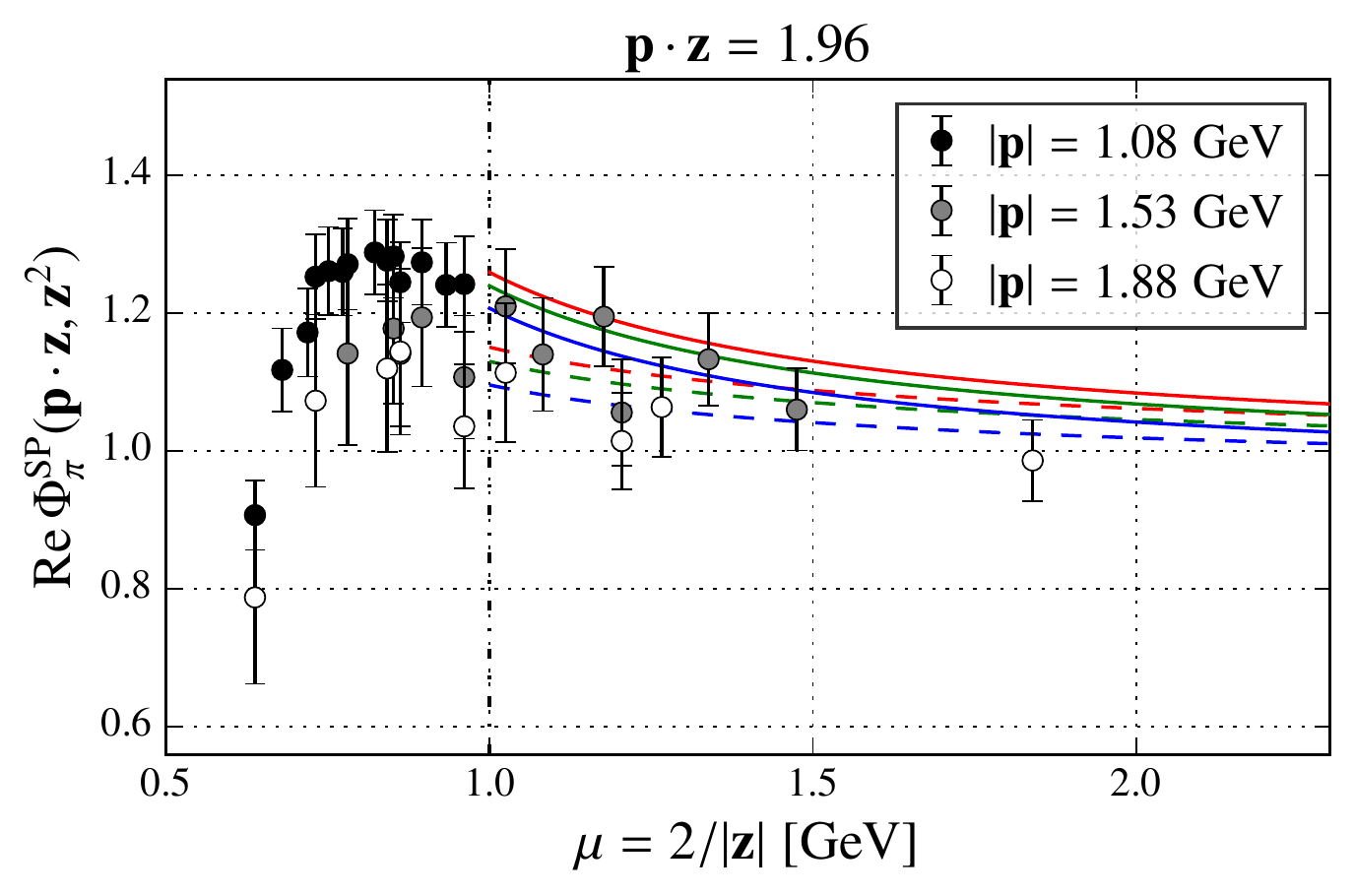}%
\caption{\label{figure_SP}The plots show our results for the scalar-pseudoscalar channel for different fixed values of $\mathbf p \cdot \mathbf z$ and different pion momenta. The errorbars include the statistical error only. The solid/dashed lines correspond to predictions taking into account/neglecting higher twist contributions for the different DA models~\eqref{models}. The color-coding is the same as in Fig.~\protect\ref{fig_pionDA}.}%
\end{figure*}%
In the continuum, the \emph{chiral even} part of the propagator connecting the two local currents (proportional to $\slashed z$) gives the most important contribution, while the \emph{chiral odd} part (proportional to the unit matrix) is suppressed by a factor $m \sqrt{|z^2|}$ and, thus, can be set to zero in a first approximation. However, with Wilson fermions the situation is completely different. We find that the contribution from the \emph{chiral odd} part, which removes the doublers and breaks chiral symmetry, can be of the same order of magnitude as the leading contribution, cf.\ Fig.~\ref{fig_disc}. The ``jumping'' of the points nicely demonstrates the strong dependence of the lattice artifacts on the chosen direction. In particular the points along the axes [e.g., $(1,0,0)$] exhibit the largest discretization effects, while the points along the diagonal [e.g., $(1,1,1)$] are much better behaved. The large contribution of the \emph{chiral odd} part of the propagator is a peculiarity of using Wilson fermions. However, the appearance of large discretization effects is probably a general feature of all coordinate space methods.\par%
The appearance of large contributions from the chiral odd part of the propagator would lead to huge lattice artifacts in the correlator. However, the perturbative calculation shows that in the situation where the two currents are located symmetrically with respect to the chosen origin, contributions from the chiral even and the chiral odd parts of the propagator are nicely separated in some channels. For instance, for the scalar-pseudoscalar channel the contribution from the chiral even part (which is the one we are interested in) is real, while the chiral odd part appears only in the imaginary part. Hence, we can choose to analyze only the part of the signal that does not contain the problematic contributions. Note, however, that the continuum expectation that either the real or the imaginary part (depending on which one corresponds to the chiral odd part) of the signal should be strongly suppressed is \emph{not valid} for the lattice data. Therefore, the correct identification of the relevant part of the signal is crucial.\par%
We can now concentrate on the correction of the discretization effects in the chiral even part of the propagator (these correspond to the blue points in Fig.~\ref{fig_disc}). First and foremost we have decided to simply discard data points where the free field discretization effect is already larger than $10\%$, which mainly excludes very small distances ($|\mathbf z|\lesssim 2a$) and directions along the lattice axes. For the remaining data points we use a correction factor $c^\text{corr}(z)$ determined such that the corrected propagator%
\begin{align}
 G^{\text{corr}}_{\text{latt}}(z)&\equiv c^\text{corr}(z) G_{\text{latt}}(z) \,,
\end{align}%
satisfies the condition
\begin{align}
 \operatorname{tr}_{cD}\bigl\{\slashed z G^{\text{corr}}_{\text{latt}}(z)\bigr\} &\overset{!}{=} \operatorname{tr}_{cD}\bigl\{\slashed z G_{\text{cont}}(z) \bigr\} \,, \label{improvement_condition}
\end{align}%
where the trace runs over Dirac and color indices. To zeroth order accuracy in $\alpha_s$ (where $G^{\phantom{\mathrlap{\text{free}}}}_{\text{latt}}=G^{\text{free}}_{\text{latt}}$ is the free propagator) this leads to%
\begin{align}
 c^{\text{corr}}(z) &= \Biggl(\!\operatorname{tr}_D\bigl\{\slashed z G^{\text{free}}_{\text{latt}}(z)\bigr\} \frac{z^2 \pi^2}{2}\!\Biggr)^{-1} \!\! \frac{-m^2 z^2}{2} K_2\Bigl(m\textstyle\sqrt{-z^2}\Bigr)  \,,
\end{align}%
which corresponds to multiplying the blue data points in Fig.~\ref{fig_disc} with a factor such that one obtains the continuum result in the non-interacting case. Looking at Fig.~\ref{fig_disc_corr} it is clear to the naked eye that this correction leads to a much smoother and less direction-dependent behavior of the data points.%
\section{Results}\label{sect_results}%
\begin{figure*}[p]%
\centering%
\includegraphics[width=0.44\textwidth]{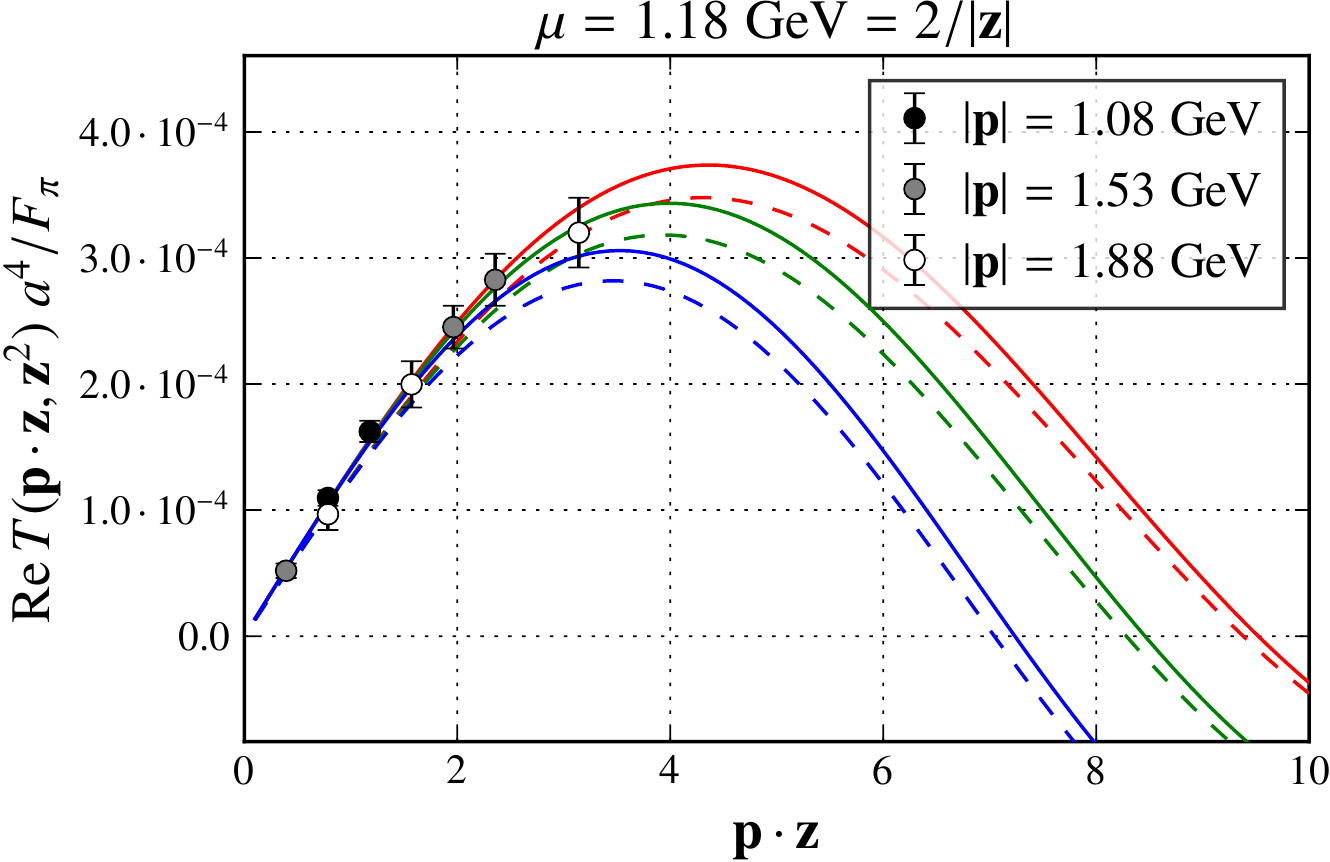}\hspace{0.02\textwidth}%
\includegraphics[width=0.44\textwidth]{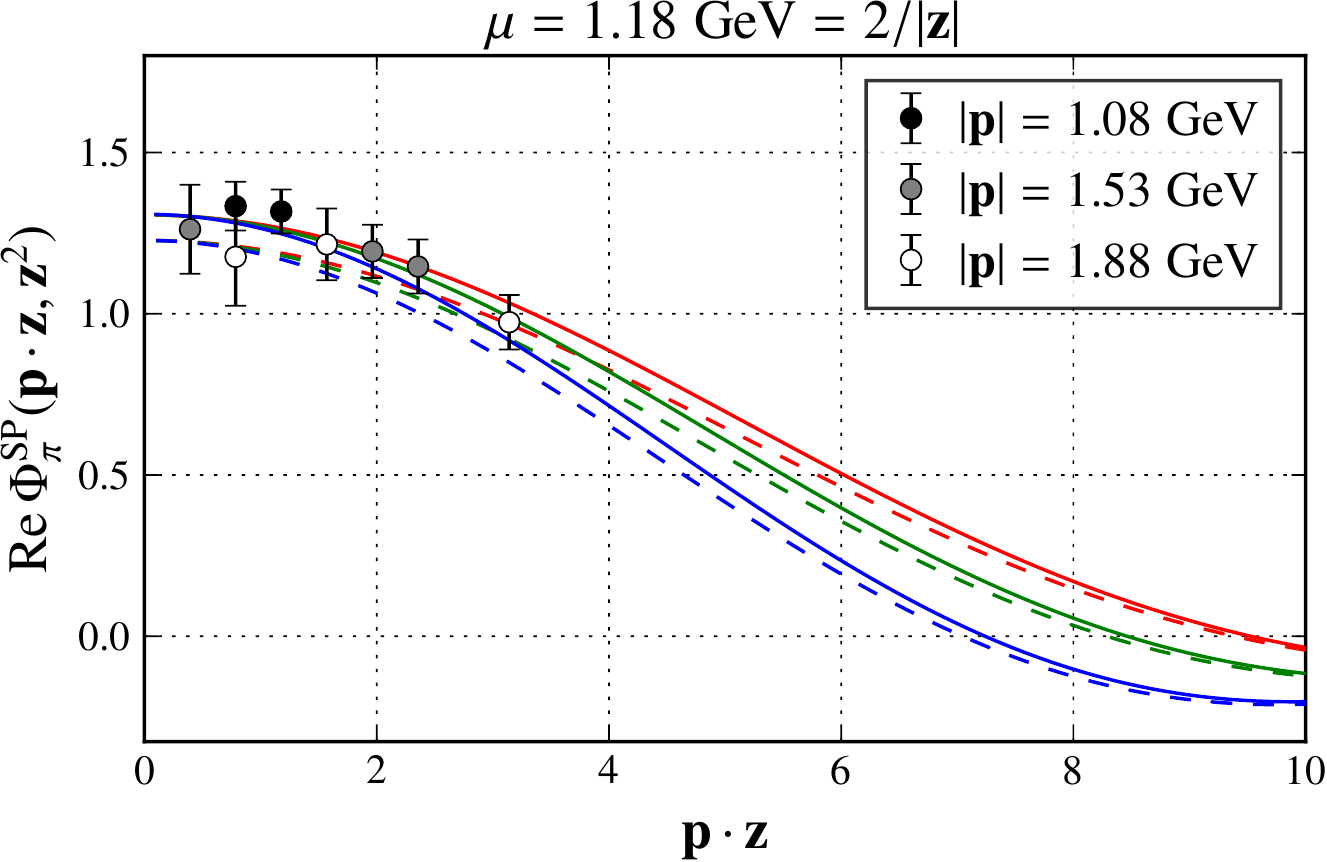}\\[0.45\baselineskip]%
\includegraphics[width=0.44\textwidth]{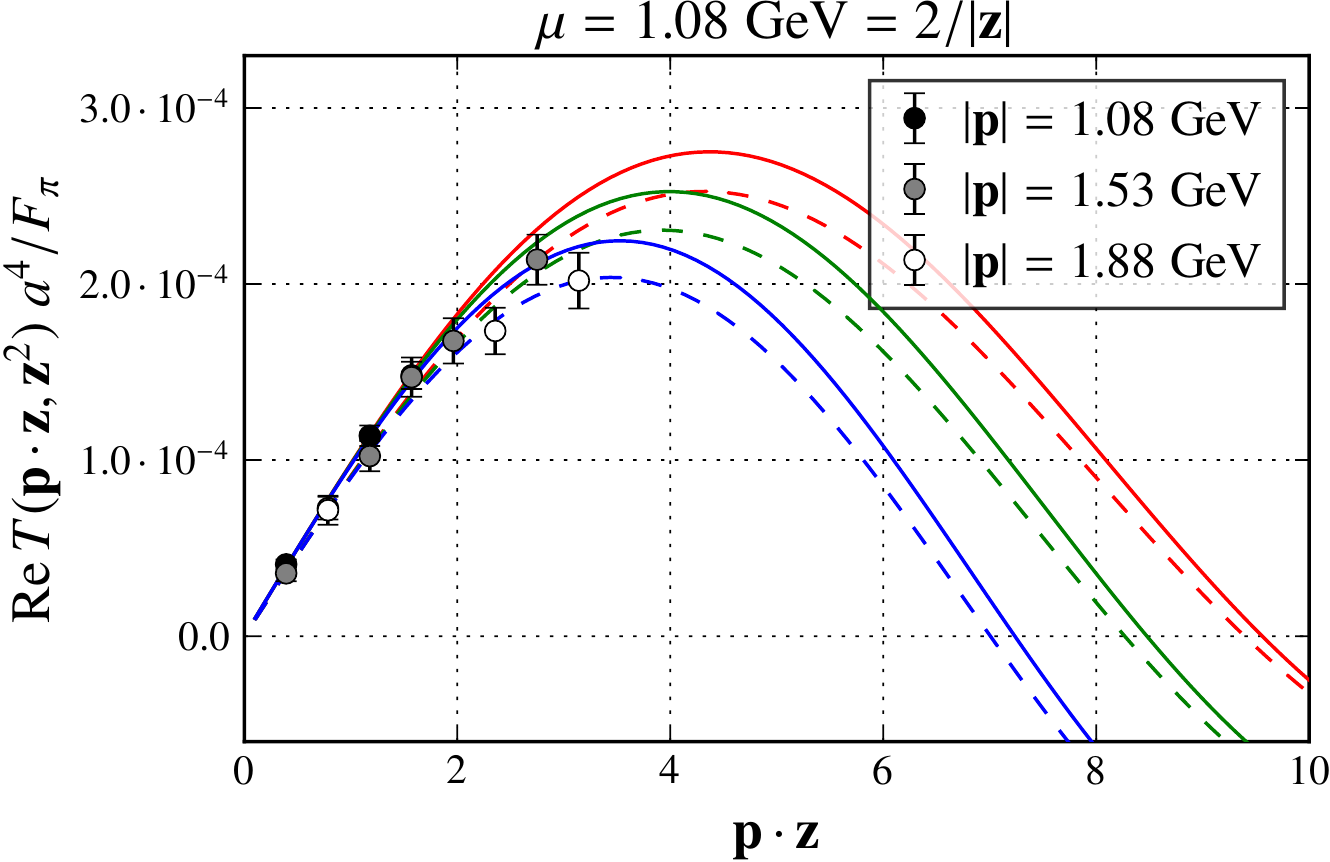}\hspace{0.02\textwidth}%
\includegraphics[width=0.44\textwidth]{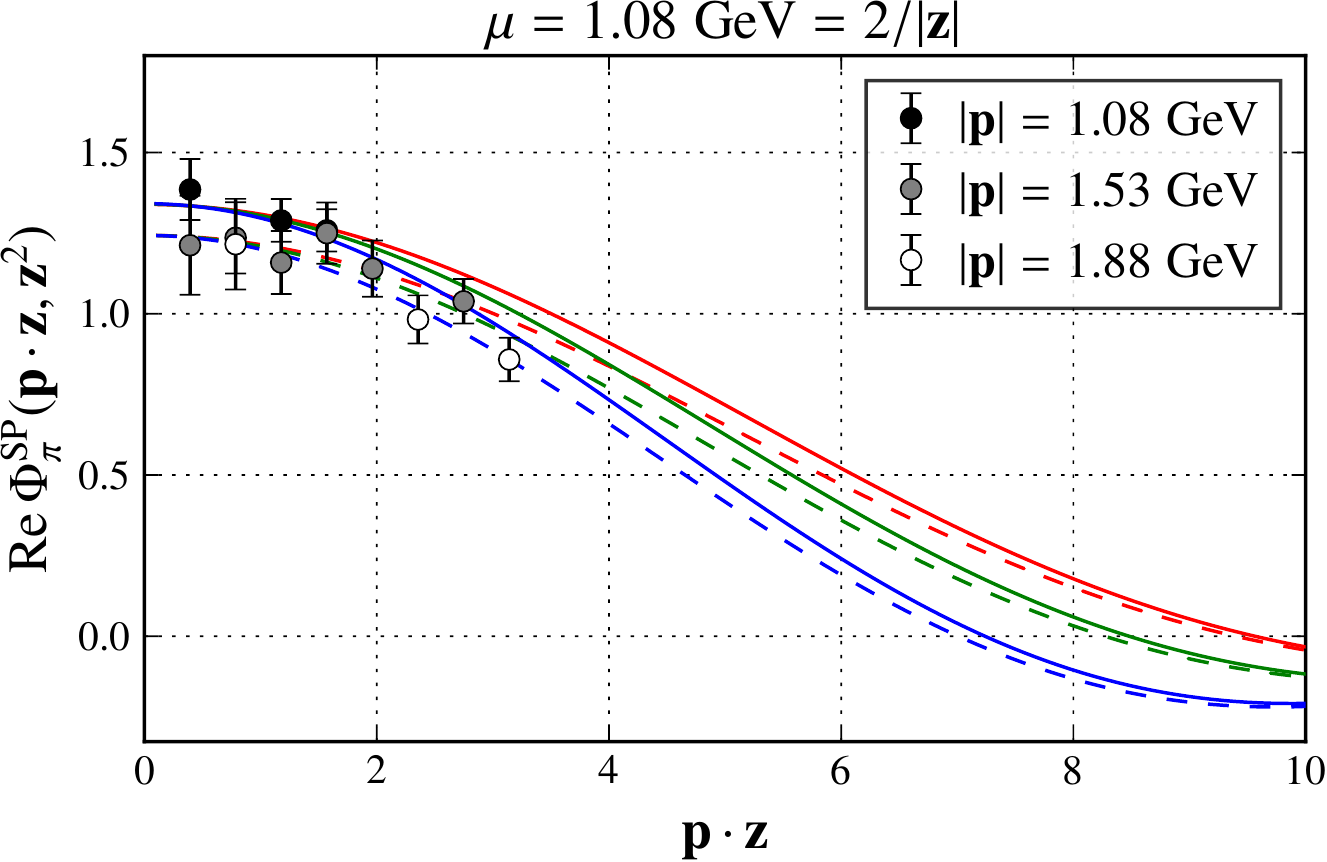}\\[0.45\baselineskip]%
\includegraphics[width=0.44\textwidth]{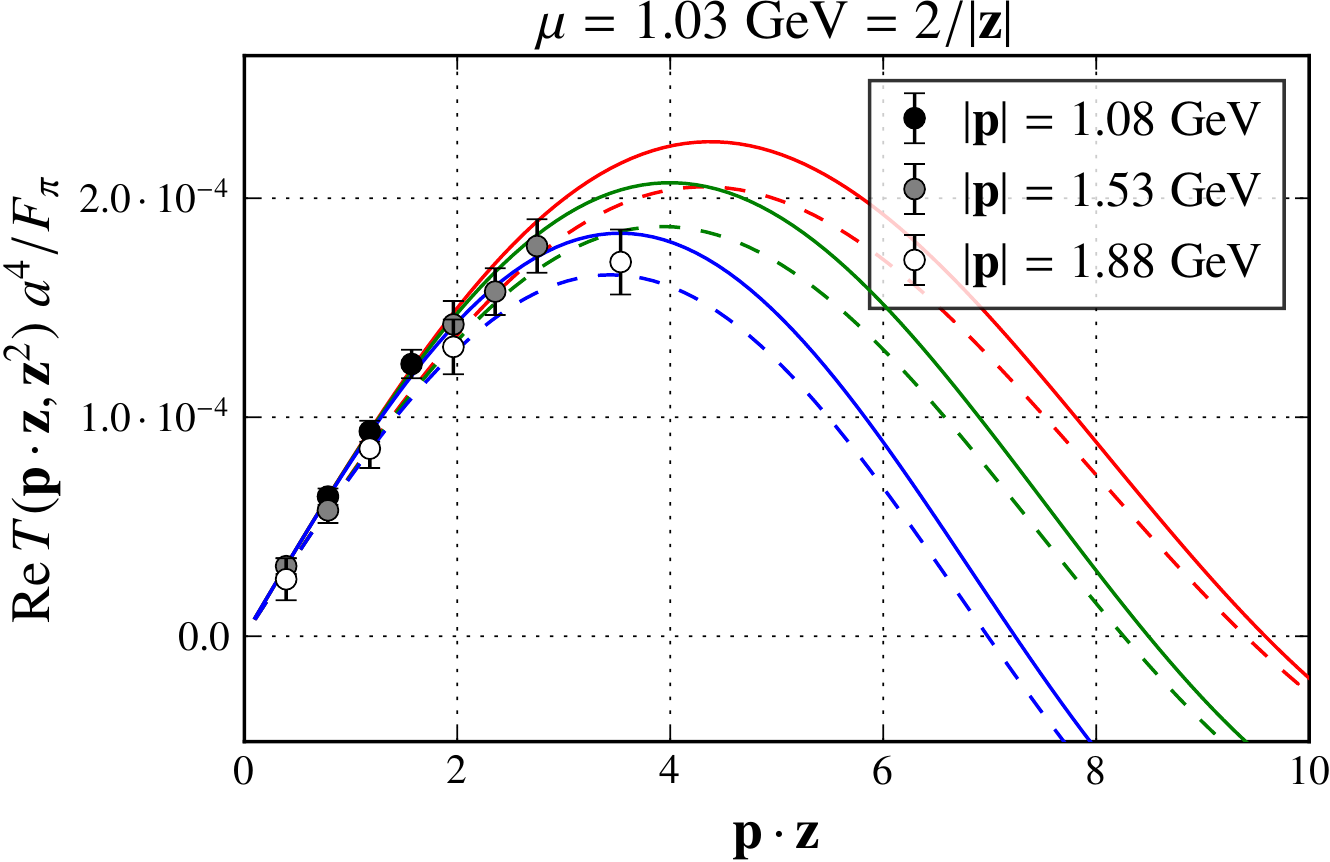}\hspace{0.02\textwidth}%
\includegraphics[width=0.44\textwidth]{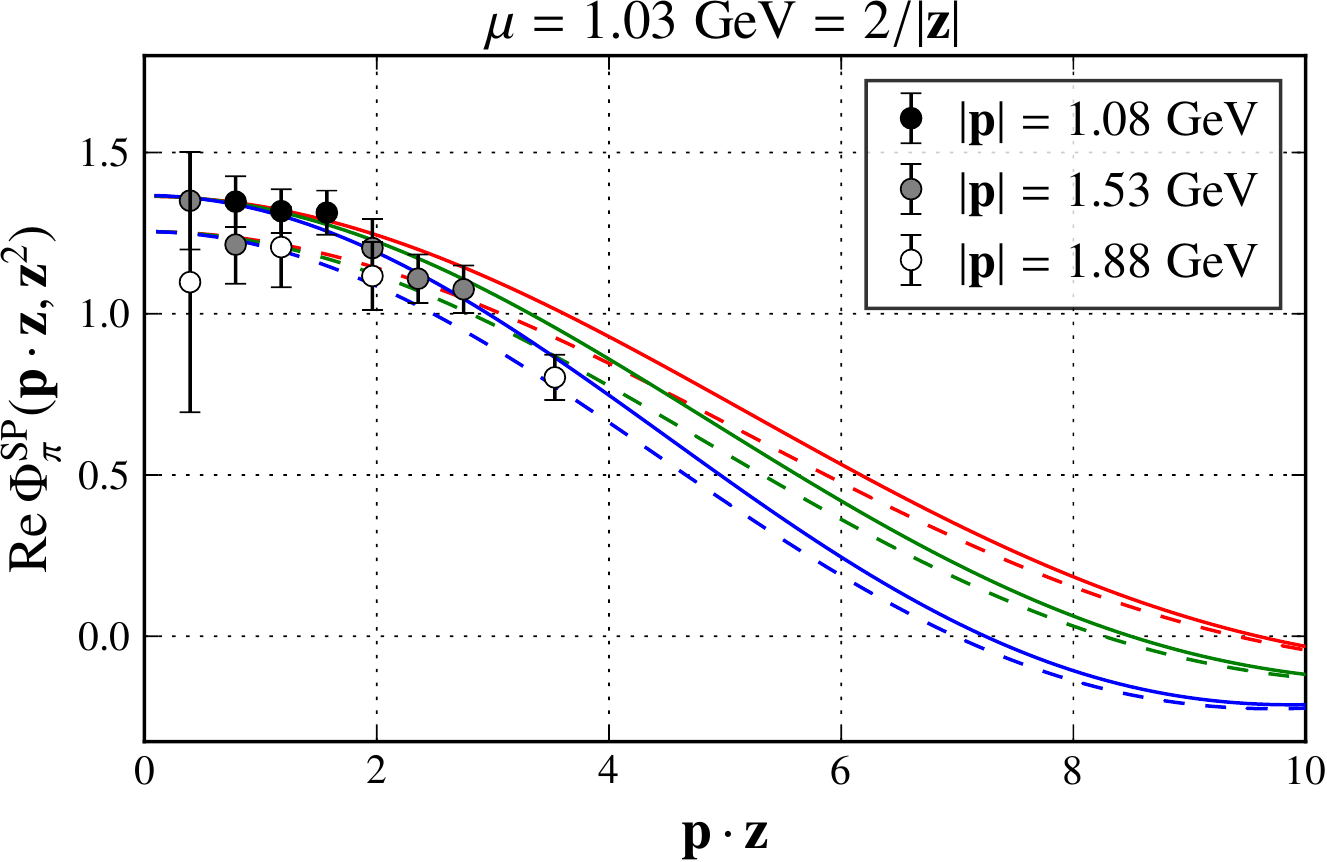}\\[0.45\baselineskip]
\includegraphics[width=0.44\textwidth]{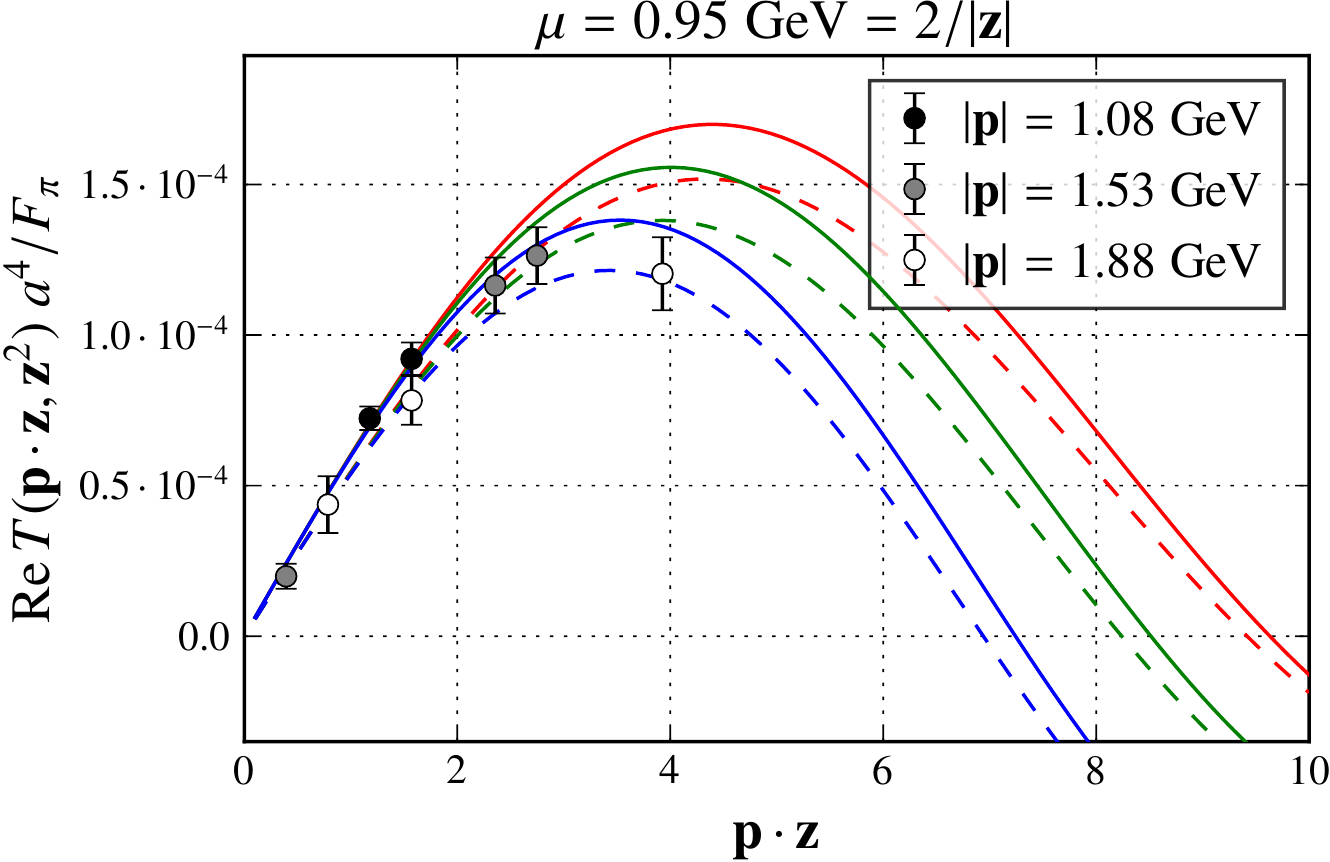}\hspace{0.02\textwidth}%
\includegraphics[width=0.44\textwidth]{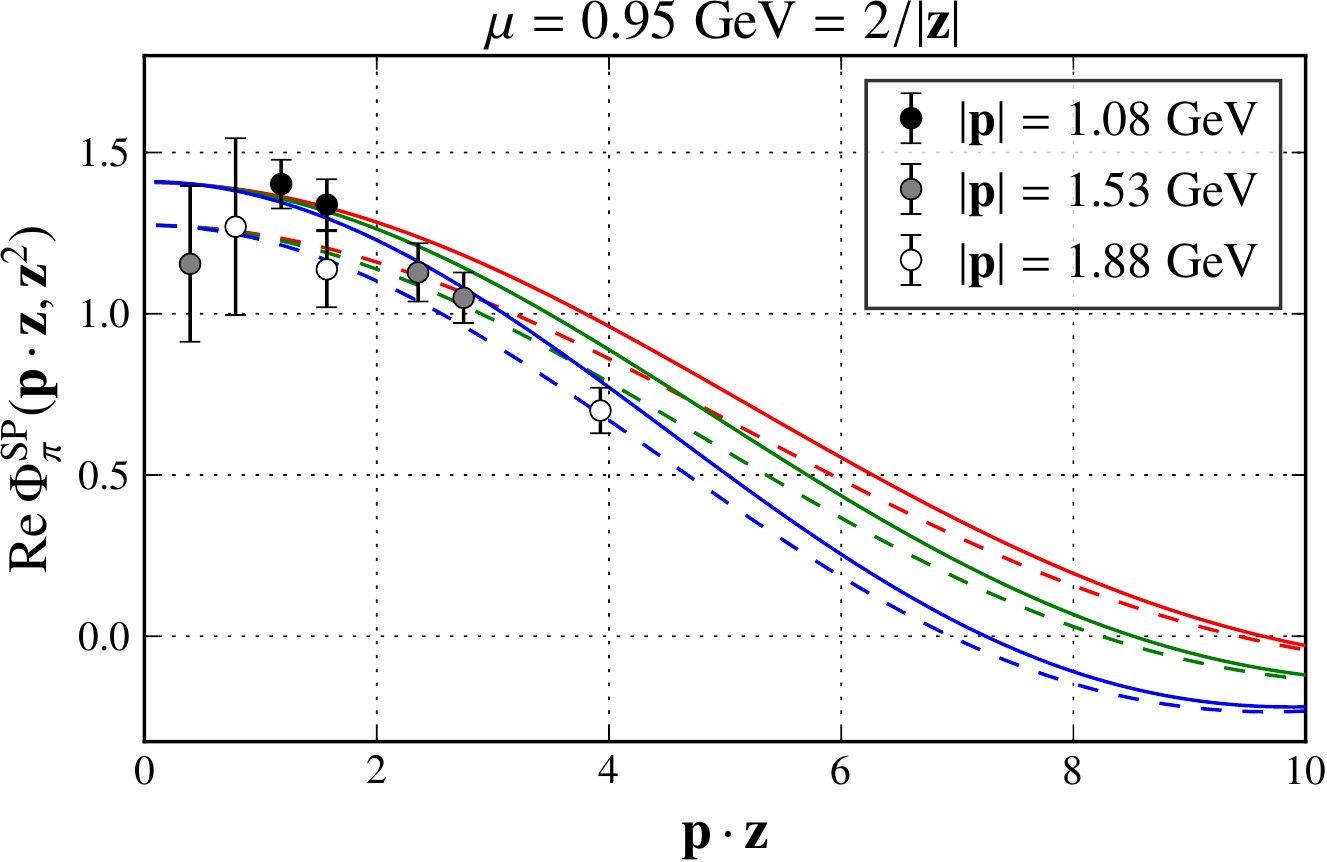}%
\caption{\label{figure_SPpx}Results from the scalar-pseudoscalar channel for fixed values of the distance $\mathbf z^2$ corresponding to perturbative scales around $\mu\approx\unit{1}{\giga\electronvolt}$. The errorbars only include the statistical error. Note that this comprises only a small subset of the available data. The left-hand-side plots display the real part of the normalized matrix element $T(\mathbf p \cdot \mathbf z,\mathbf z^2)/F_\pi$, while those on the right-hand-side show the respective real part of $\Phi_\pi^{\rm SP}(\mathbf p \cdot \mathbf z,\mathbf z^2)$ defined in Eq.~\eqref{T_SP_LO}. At tree level and up to higher twist corrections, the latter correspond to the position space DA~\eqref{Phi_pi}. Solid/dashed lines correspond to predictions that include/neglect higher twist contributions. The color-coding is the same as in figure~\protect\ref{fig_pionDA}.}%
\end{figure*}%
The gauge field ensemble used in this study has been generated (by QCDSF / RQCD) with two mass-degenerate flavors of nonperturbatively improved Wilson fermions and the Wilson gluon action (ensemble~IV in Ref.~\cite{Bali:2014gha}). The dimensions of the lattice are $32^3\times64$ and the hopping parameter is $\kappa=0.13632$. The coupling parameter $\beta=5.29$ translates to the lattice spacing $a\approx\unit{0.071}{\femto\meter}=(\unit{2.76}{\giga\electronvolt})^{-1}$ and the pion mass has been determined in Ref.~\cite{Bali:2014nma} to the value $m_\pi=0.10675(59)/a \approx \unit{295}{\mega\electronvolt}$. In order to get a reasonable overlap with the hadron state at large momentum, we have employed the momentum smearing technique (cf.\ Ref.~\cite{Bali:2016lva}) with APE smeared links \cite{Falcioni:1984ei}.\par%
The operator renormalization is performed as described in Ref.~\cite{Gockeler:2010yr}. The local operators are renormalized nonperturbatively in a RI${}^\prime$-MOM scheme along with a subtraction of lattice artifacts in one-loop perturbation theory. The final conversion to the $\MSbar$ scheme employs $3$ loop continuum perturbation theory. To be consistent, we use the $N_f=2$ specific running of $\alpha_s$ in all perturbative calculations. To this end, we combine the results of Refs.~\cite{Fritzsch:2012wq} and~\cite{Bali:2012qs} to obtain a value of $\alpha_s$ at $1000/a\approx\unit{2.76}{\tera\electronvolt}$. From there we evolve it downwards using $5$ loop running.\par%
In Figs.~\ref{figure_SP} and~\ref{figure_SPpx} we confront the data points with predictions from continuum perturbation theory corresponding to the pion DAs shown in Fig.~\ref{fig_pionDA}. For all cases we show a version ignoring higher twist (i.e., twist~$4$) effects (dashed lines) and one including higher twist corrections (solid lines), where $\delta^\pi_2(\unit{1}{\giga\electronvolt})=\unit{0.2}{\giga\electronvolt\squared}$ is set to the QCD sum rule estimate obtained in Ref.~\cite{Novikov:1983jt} (cf.\ also Ref.~\cite{Braun:1999uj}). The errorbars only include the statistical error. It is clear that the systematic uncertainty is sizeable: in addition to discretization effects and higher order perturbative corrections there may be excited state contaminations and, possibly, finite volume effects. Therefore, one should refrain from drawing any premature phenomenological conclusions. Nevertheless, the qualitative agreement found in Ref.~\cite{Chen:2017gck} between our data points at $\mu=\unit{1.08}{\giga\electronvolt}$ and the results obtained using the quasi-DA method is encouraging.\par%
In Fig.~\ref{figure_SP}, one immediately notices the higher twist effect, in particular for large distances, while the curves corresponding to the various DAs are hard to distinguish for small values of $\mathbf p \cdot \mathbf z$. In Fig.~\ref{figure_SPpx} deviations from the asymptotic form are nicely visible at $\mathbf p \cdot \mathbf z \gtrsim 4$. The latter region, however, can only be reached with larger hadron momenta, since we are limited to perturbatively accessible values of $|\mathbf z|=2/\mu \lesssim \unit{2}{\giga\electronvolt^{-1}}\approx 5.5 a$. In Fig.~\ref{figure_SPpx} it becomes clear that our data points with $|\mathbf p|=\unit{1.88}{\giga\electronvolt}$ can already reach out into this region, but that one still needs higher statistics to be able to differentiate between different DA models.%
\section{Summary}%
In this work we have demonstrated that the coordinate space method for the determination of the pion DA proposed in Ref.~\cite{Braun:2007wv} is promising, in particular as far as the statistical error is concerned. To this end, we have analyzed lattice data at $m_\pi=\unit{295}{\mega\electronvolt}$ at a lattice spacing of $a=\unit{0.071}{\femto\meter}$ using dynamic Wilson fermions. We have shown that the large hadron momenta, which are a prerequisite of this method (and also for other related methods), lie just within the scope of the novel momentum smearing technique~\cite{Bali:2016lva}.\par%
We have found particularly large discretization effects for directions along the coordinate axes, which are probably not specific to our calculation but will most likely occur also in other coordinate space calculations. Furthermore, we find that the chiral odd part of the quark propagator leads to large lattice artifacts. This contribution stems from the Wilson term in the propagator and is therefore a peculiarity of using Wilson fermions. We have overcome this problem by analyzing the real part of the scalar-pseudoscalar channel, where only the chiral even part contributes. For the remaining discretization effects stemming from the propagator, we have adopted the correction method described in Sect.~\ref{sect_disc_corr}, which has reduced the anisotropy of the data considerably. However, observing large discretization effects on this single intermediate lattice spacing shows that taking the continuum limit will be of vital importance, if one aims at achieving quantitative results in the future.\par%
Unlike the Wilson line approach~\cite{Zhang:2017bzy,Orginos:2017kos}, the direction of the separation between the currents $\mathbf z$ can be chosen arbitrarily using our method. This enables us to realize a large number of different $|\mathbf z|$ and $\mathbf p \cdot \mathbf z$ values and also to study and minimize discretization effects. Intricacies related to the Wilson line renormalization~\cite{Ishikawa:2016znu,Ishikawa:2017jtf,Rossi:2017muf,Ji:2017rah,Ji:2017oey,Orginos:2017kos,Alexandrou:2017huk,Green:2017xeu} are avoided entirely, and the possibility to vary the Dirac structures in the currents offers an additional handle on the higher-order perturbative corrections and higher twist effects.\par%
In the near future we plan to investigate a new algorithm that may reduce the statistical uncertainties. We will also move to a smaller lattice spacing to enable the use of larger momenta $|\mathbf p|\ll \pi/a$ (and therefore larger $\mathbf p\cdot\mathbf z$ values at a given scale $\mu=2/|\mathbf z|\gtrsim \unit{1}{\giga\electronvolt}$) along with distances $|\mathbf z|\gg a$, which will reduce lattice artifacts.\par%
\begin{acknowledgement}%
This work has been supported by the Deut\-sche For\-schungs\-ge\-mein\-schaft (SFB/TRR\nobreakdash-55) and the Stu\-di\-en\-stif\-tung des deut\-schen Vol\-kes. The analysis was carried out on the QPACE~2~\cite{Arts:2015jia} Xeon~Phi installation of the SFB/TRR\nobreakdash-55 in Regensburg. We used a modified version of the {\sc Chroma}~\cite{Edwards:2004sx} software package along with the {\sc Lib\-Hadron\-Analysis} library and the multigrid solver implementation of Ref.~\cite{Heybrock:2015kpy} (see also Ref.~\cite{Frommer:2013fsa}). We thank Daniel Richtmann for code development, discussions and software support.\par%
\end{acknowledgement}%
\raggedbottom
%

%
%
\pagebreak
%
\end{document}